\documentclass[12pt]{article}
\pdfoutput=1
\usepackage[colorlinks,linkcolor=Blue,citecolor=Blue,bookmarks,bookmarksnumbered]{hyperref}
\usepackage[scaled=0.85]{helvet}
\usepackage{amsmath,amssymb,accents,mathrsfs,XoohmE}
\usepackage{graphicx,color}
\usepackage{booktabs}
\usepackage{multirow}
\usepackage{placeins}
\usepackage{amsmath}

\definecolor{Green}  {rgb}{0.10,0.70,0.10} 
\definecolor{Orange} {rgb}{1.00,0.50,0.15} 
\definecolor{Red}    {rgb}{0.90,0.00,0.12} 
\definecolor{Purple} {rgb}{0.50,0.25,0.55} 
\definecolor{Turque} {rgb}{0.00,0.65,0.85} 
\definecolor{Blue}   {rgb}{0.00,0.00,1.00} 
\definecolor{Magenta}{rgb}{1.00,0.00,1.00} 
\definecolor{Gold}   {rgb}{1.00,0.75,0.25} 
\definecolor{Seaweed}{rgb}{0.01,0.24,0.09} 
\definecolor{Brown}  {rgb}{0.43,0.26,0.32} 
\definecolor{grey1}  {rgb}{0.20,0.20,0.20} 
\definecolor{grey2}  {rgb}{0.40,0.40,0.40} 
\definecolor{grey3}  {rgb}{0.60,0.60,0.60} 
\definecolor{grey4}  {rgb}{0.80,0.80,0.80} 
\definecolor{grey5}  {rgb}{0.90,0.90,0.90} 
\def\C#1#2{{\ifcase#1\or
             \color{Green}\or \color{Orange}\or \color{Red}\or
              \color{Purple}\or \color{Turque}\or \color{Blue}\or
               \color{Magenta}\or \color{Gold}\or \color{Seaweed}\or
                \color{Brown}\or\color{grey1}\or\color{grey2}\or
                 \color{grey3}\else\color{grey4}\fi#2}}

\definecolor{Slate} {rgb}{0.00,0.45,0.55}

\definecolor{AdinkraGreen}{rgb}{0.10196079, 0.61176473, 0.21960784 }
\definecolor{AdinkraViolet}{rgb}{0.42352942, 0.15294118, 0.4509804 }
\definecolor{AdinkraOrange}{rgb}{0.89803922, 0.57647061, 0.27450982}
\definecolor{AdinkraRed}{rgb}{0.78431374, 0, 0.12156863}

\def\rI{{\rm I}}
\def\rJ{{\rm J}}
\def\rK{{\rm K}}
\def\rL{{\rm L}}

\def\vCent#1{\vcenter{\hbox{\hss#1\hss}}}
\def\be{\begin{equation}}
\def\ee{\end{equation}}
\newcommand{\bea}{\begin{eqnarray}}
\newcommand{\eea}{\end{eqnarray}}
\newcommand{\ena}{\end{eqnarray}}

\def\pp{{\mathchoice
          {
              \kern 1pt%
              \raise 1pt
              \vbox{\hrule width5pt height0.4pt depth0pt
                    \kern -2pt
                    \hbox{\kern 2.3pt
                          \vrule width0.4pt height6pt depth0pt
                          }
                    \kern -2pt
                    \hrule width5pt height0.4pt depth0pt}%
                    \kern 1pt
           }
            {
              \kern 1pt%
              \raise 1pt
              \vbox{\hrule width4.3pt height0.4pt depth0pt
                    \kern -1.8pt
                    \hbox{\kern 1.95pt
                          \vrule width0.4pt height5.4pt depth0pt
                          }
                    \kern -1.8pt
                    \hrule width4.3pt height0.4pt depth0pt}%
                    \kern 1pt
            }
            {
              \kern 0.5pt%
              \raise 1pt
              \vbox{\hrule width4.0pt height0.3pt depth0pt
                    \kern -1.9pt  
                    \hbox{\kern 1.85pt
                          \vrule width0.3pt height5.7pt depth0pt
                          }
                    \kern -1.9pt
                    \hrule width4.0pt height0.3pt depth0pt}%
                    \kern 0.5pt
            }
            {
              \kern 0.5pt%
              \raise 1pt
              \vbox{\hrule width3.6pt height0.3pt depth0pt
                    \kern -1.5pt
                    \hbox{\kern 1.65pt
                          \vrule width0.3pt height4.5pt depth0pt
                          }
                    \kern -1.5pt
                    \hrule width3.6pt height0.3pt depth0pt}%
                    \kern 0.5pt
            }
        }}

\def\mm{{\mathchoice
                       {
                             \kern 1pt
               \raise 1pt    \vbox{\hrule width5pt height0.4pt depth0pt
                                  \kern 2pt
                                  \hrule width5pt height0.4pt depth0pt}
                             \kern 1pt}
                       {
                            \kern 1pt
               \raise 1pt \vbox{\hrule width4.3pt height0.4pt depth0pt
                                  \kern 1.8pt
                                  \hrule width4.3pt height0.4pt depth0pt}
                             \kern 1pt}
                       {
                            \kern 0.5pt
               \raise 1pt
                            \vbox{\hrule width4.0pt height0.3pt depth0pt
                                  \kern 1.9pt
                                  \hrule width4.0pt height0.3pt depth0pt}
                            \kern 1pt}
                       {
                           \kern 0.5pt
             \raise 1pt  \vbox{\hrule width3.6pt height0.3pt depth0pt
                                  \kern 1.5pt
                                  \hrule width3.6pt height0.3pt depth0pt}
                           \kern 0.5pt}
                       }}

\def\ad{{\kern0.5pt
                   \alpha \kern-5.05pt \raise5.8pt\hbox{$\textstyle.$}\kern
0.5pt}}

\def\bd{{\kern0.5pt
                   \beta \kern-5.05pt \raise5.8pt\hbox{$\textstyle.$}\kern
0.5pt}}

\def\qd{{\kern0.5pt
                   q \kern-5.05pt \raise5.8pt\hbox{$\textstyle.$}\kern
0.5pt}}
\def\Dot#1{{\kern0.5pt
     {#1} \kern-5.05pt \raise5.8pt\hbox{$\textstyle.$}\kern
0.5pt}}

\catcode`@=11
\def\un#1{\relax\ifmmode\@@underline#1\else
        $\@@underline{\hbox{#1}}$\relax\fi}
\catcode`@=12

\def\a{\alpha}
\def\b{\beta}
\def\c{\chi}
\def\d{\delta}
\def\e{\epsilon}

\def\g{\gamma}

\def\i{\iota}
\def\j{\psi}
\def\k{\kappa}
\def\l{\lambda}
\def\m{\mu}
\def\n{\nu}
\def\o{\omega}

\def\r{\rho}
\def\s{\sigma}
\def\t{\tau}

\def\D{\Delta}

\def\L{\Lambda}
\def\O{\Omega}
\def\P{\Pi}

\def\S{\Sigma}

\def\ca{{\cal A}}

\def\cf{{\cal F}}

\def\dslash{\not{\hbox{\kern-2pt $\partial$}}}
\def\Dslash{\not{\hbox{\kern-4pt $D$}}}
\def\pslash{\not{\hbox{\kern-2.3pt $p$}}}
 \newtoks\slashfraction
 \slashfraction={.13}
 \def\slash#1{\setbox0\hbox{$ #1 $}
 \setbox0\hbox to \the\slashfraction\wd0{\hss \box0}/\box0 }
 
\def\Sc#1{{\hbox{\sc #1}}}      
\def\kcr{{\hbox{\ro \char'170}}}                
\def\ktl{{\hbox{\ro \char'170}}}        
\def\ktr{{\hbox{\ro \char'170}}}        
\def\kbl{{\hbox{\ro \char'170}}}        
\def\kbr{{\hbox{\ro \char'170}}}        

\def\plpl{\raise-2pt\hbox{$\raise3pt\hbox{$_+$}\hskip-6.67pt\raise0.0pt
\hbox{$^+$}\hskip 0.01pt$}}
\def\mimi{\raise-2pt\hbox{$\raise3pt\hbox{$_-$}\hskip-6.67pt\raise0.0pt
\hbox{$^-$}\hskip 0.01pt$}} 

\def\bo{{\raise.15ex\hbox{\large$\Box$}}}               
\def\TH{{\raise.2ex\hbox{$\displaystyle \bigodot$}\mskip-4.7mu \llap H \;}}
\def\face{{\raise.2ex\hbox{$\displaystyle \bigodot$}\mskip-2.2mu \llap {$\ddot
        \smile$}}}                                      

\def\dt#1{\on{\hbox{\bf .}}{#1}}                
\def\Dot#1{\dt{#1}}

\def\Tilde#1{\widetilde{#1}}                    
\def\Hat#1{\widehat{#1}}                        
\def\leftrightarrowfill{$\mathsurround=0pt \mathord\leftarrow \mkern-6mu
        \cleaders\hbox{$\mkern-2mu \mathord- \mkern-2mu$}\hfill
        \mkern-6mu \mathord\rightarrow$}
\def\dvec#1{\vbox{\ialign{##\crcr
        \leftrightarrowfill\crcr\noalign{\kern-1pt\nointerlineskip}
        $\hfil\displaystyle{#1}\hfil$\crcr}}}           
\def\dt#1{{\buildrel {\hbox{\LARGE .}} \over {#1}}}     

\def\sfrac#1#2{{\vphantom1\smash{\lower.5ex\hbox{\small$#1$}}\over
        \vphantom1\smash{\raise.4ex\hbox{\small$#2$}}}} 
\def\bfrac#1#2{{\vphantom1\smash{\lower.5ex\hbox{$#1$}}\over
        \vphantom1\smash{\raise.3ex\hbox{$#2$}}}}       
\def\afrac#1#2{{\vphantom1\smash{\lower.5ex\hbox{$#1$}}\over#2}}    

\let\bm\relax
\newcommand{\bm}[1]{{\boldsymbol{#1}}}

\def\ad{{\dot{\alpha}}}
\def\bd{{\dot{\beta}}}

 \font\rOpe=cmsy10                        
 \def\ktl{{\hbox{\rOpe\char'170}}}        
 \def\kbl{{\hbox{\rOpe\char'170}}}        
 \def\kcr{{\reflectbox{\rOpe\char'170}}}        
 \def\ktr{{\reflectbox{\rOpe\char'170}}}        
 \def\kbr{{\reflectbox{\rOpe\char'170}}}        
 \def\Border{\vbox{\hsize0pt
        \setlength{\unitlength}{1mm}
        \newcount\xco
        \newcount\yco
        \xco=-21
        \yco=12
        \begin{picture}(0,0)(-7.5,0)
        \put(\xco,\yco){$\ktl$}
        \advance\yco by-1
        {\loop
        \put(\xco,\yco){$\kcr$}
        \advance\yco by-2
        \ifnum\yco>-240
        \repeat
        \put(\xco,\yco){$\kbl$}}
        \xco=170
        \yco=12
        \put(\xco,\yco){$\ktr$}
        \advance\yco by-1
        {\loop
        \put(\xco,\yco){$\kcr$}
        \advance\yco by-2
        \ifnum\yco>-240
        \repeat
        \put(\xco,\yco){$\kbr$}}
        \put(-19.5,13){\scalebox{.6065}{%
         University of Maryland Center for String and Particle  Theory \&\ Physics Department%
        |University of Maryland Center for String and Particle  Theory \&\ Physics Department}}
        \put(-19.5,-241.5){\scalebox{.5835}{%
         ****University of Maryland * Center for String and
         Particle  Theory* Physics Department****University of Maryland *Center
        for String and Particle  Theory* Physics Department}}
        \end{picture}
        \par\vskip-8mm}}
\definecolor{UMred}{rgb}{.9,.05,.2}
\definecolor{HUblue}{rgb}{.0,.3,.7}
 \def\UMbanner{\vbox{\hsize0pt
        \setlength{\unitlength}{.4mm}
        \thicklines  
        \begin{picture}(0,0)(-30,-10)
        \put(165,2){\line(1,0){4}}
        \put(170,2){\line(1,0){4}}
        \put(180,2){\line(1,0){4}}
        \put(175,-14){\line(1,0){4}}
        \put(180,-14){\line(1,0){4}}
        \put(185,-14){\line(1,0){4}}
        \put(169,-14){\line(0,1){16}}
        \put(170,-14){\line(0,1){16}}
        \put(179,-14){\line(0,1){16}}
        \put(180,-14){\line(0,1){16}}
        \put(184,-14){\line(0,1){16}}
        \put(185,-14){\line(0,1){16}}
        \put(169,2){\oval(8,32)[bl]}
        \put(170,2){\oval(8,32)[br]}
        \put(179,-14){\oval(8,32)[tl]}
        \put(185,-14){\oval(8,32)[tr]}
        \end{picture}
        \par\vskip-6.5mm
        \thicklines}}

\definecolor{Red}    {rgb}{0.90,0.00,0.12} 
\definecolor{Blue}   {rgb}{0.00,0.00,1.00} 
\definecolor{Green}  {rgb}{0.10,0.70,0.10} 
\definecolor{Turque} {rgb}{0.00,0.65,0.85} 
\definecolor{Orange} {rgb}{1.00,0.50,0.15} 
\definecolor{Magenta}{rgb}{1.00,0.00,1.00} 
\definecolor{Gold}   {rgb}{1.00,0.75,0.25} 
\definecolor{Seaweed}{rgb}{0.01,0.24,0.09} 
\definecolor{Purple} {rgb}{0.50,0.25,0.55} 
\definecolor{Brown}  {rgb}{0.43,0.26,0.32} 
\definecolor{grey1}  {rgb}{0.20,0.20,0.20} 
\definecolor{grey2}  {rgb}{0.40,0.40,0.40} 
\definecolor{grey3}  {rgb}{0.60,0.60,0.60} 
\definecolor{grey4}  {rgb}{0.80,0.80,0.80} 
\definecolor{grey5}  {rgb}{0.90,0.90,0.90} 
\def\C#1#2{{\ifcase#1\or
             \color{Red}\or \color{Green}\or \color{Blue}\or\
              \color{Turque}\or \color{Orange}\or \color{Magenta}\or 
               \color{Gold}\or \color{Seaweed}\or \color{Purple}\or
                \color{Brown}\or\color{grey1}\or\color{grey2}\or
                 \color{grey3}\else\color{grey4}\fi#2}}

\definecolor{Slate} {rgb}{0.00,0.45,0.55}

\newdimen\parshift\parshift=\parindent
\catcode`@=11
 \long\def\@footnotetext#1{\insert\footins{\reset@font\footnotesize
           \interlinepenalty\interfootnotelinepenalty\splittopskip%
            \footnotesep\splitmaxdepth\dp\strutbox\floatingpenalty\@MM%
             \hsize\columnwidth\addtolength{\hsize}{-2\parindent}
              \@parboxrestore\protected@edef\@currentlabel%
              {\csname p@footnote\endcsname\@thefnmark}%
                \color@begingroup%
                 \@makefntext{\rule\z@\footnotesep\ignorespaces#1%
                  \@finalstrut\strutbox}%
                \color@endgroup}}
 \long\def\@makefntext#1{\hglue\parshift%
           \vbox{\noindent\baselineskip=11pt plus.5pt minus.5pt\hb@xt@0em{\hss\@makefnmark\kern1pt}#1}}
\catcode`@=12

\newskip\humongous \humongous=0pt plus 1000pt minus 1000pt
\def\caja{\mathsurround=0pt}
\def\eqalign#1{\,\vcenter{\openup2\jot \caja
        \ialign{\strut \hfil$\displaystyle{##}$&$
        \displaystyle{{}##}$\hfil\crcr#1\crcr}}\,}
\newif\ifdtup
\def\panorama{\global\dtuptrue \openup2\jot \caja
        \everycr{\noalign{\ifdtup \global\dtupfalse
        \vskip-\lineskiplimit \vskip\normallineskiplimit
        \else \penalty\interdisplaylinepenalty \fi}}}

\makeatletter
\def\section{\@startsection{section}{1}{\z@}
        {3ex plus-1ex minus-.2ex}{1pt plus1pt}{\large\sf\bfseries\boldmath}}
\def\subsection{\@startsection{subsection}{2}{\z@}
         {1.5ex plus-1ex minus-.2ex}{0.01pt plus1pt}{\sf\slshape}}
\def\subsubsection{\@startsection{subsubsection}{3}{\z@}
          {1.5ex plus-1ex minus-.2ex}{0.01pt plus0.2pt}{\sf\boldmath}}
\def\paragraph{\@startsection{paragraph}{4}{\z@}
           {.75ex \@plus.5ex \@minus.2ex}{-2mm}{\sf\bfseries\boldmath}}
\makeatother

 \allowdisplaybreaks
 \seceq

\def\DDt#1{\accentset{\hbox{\LARGE.\kern-2pt.}}{#1}}	
\def\dt#1{\accentset{\hbox{\large.}}{#1}}	
\def\ddt#1{\accentset{\hbox{\large\kern.5pt.\kern-1pt.}}{#1}}	
\def\eX{\rlap{\raisebox{.35ex}{\kern.45ex\scriptsize\it=}}{\boldsymbol X}}
\def\eY{\rlap{\raisebox{.35ex}{\kern.12ex\scriptsize\it=}}{\boldsymbol Y}}
\def\EX{\rlap{\raisebox{.45ex}{\kern.425ex\scriptsize\it=}}{X}}
\def\EY{\rlap{\raisebox{.45ex}{\kern.1ex\scriptsize\it=}}{Y}}

\def\ad{{\dt\a}}
\def\bd{{\dt\b}}

\def\bDb{\hbox{\kern2pt\vrule height10pt depth-9.2pt width6pt\kern-9pt{$\boldsymbol D$}}\mkern-2mu}

\def\bQb{\hbox{\kern2pt\vrule height10pt depth-9.2pt width6pt\kern-9pt{$\boldsymbol Q$}}}

\def\BSb{\hbox{\kern2.5pt\vrule height10pt depth-9.2pt width7pt\kern-10.25pt{$\boldsymbol{\mit\Sigma}$}}}

\def\rQb{\hbox{\kern1pt\vrule height10pt depth-9.2pt width6pt\kern-8pt{\bf Q}}}

\def\rBx#1#2{\hbox to#1{#2\hss}}

\begin{document}

\thispagestyle{empty}
\vbox{\Border\UMbanner}
\noindent{\small
\today\hfill{PP-017-019 \\ 
$~~~~~~~~~~~~~~~~~~~~~~~~~~~~~~~~~~~~~~~~~~~~~~~~~~~~~~~~~~~~$
$~~~~~~~~~~~~~~~~~~~~\,~~~~~~~~~~~~~~~~~~~~~~~~~\,~~~~~~~~~~~~~~~~$
 {HET-1709}
}
\vspace*{4mm}
\begin{center}
{\large \bf
Spacetime Spin and Chirality Operators for 
\\[6pt]  
Minimal
4D, $\cal N$ = 1
Supermultiplets From 
\\[6pt]  
BC$\bm{{}_4}$
Adinkra-Tessellation of Riemann Surfaces}   \\   [6mm]
{\large {
S.\ James Gates, Jr.,\footnote{gatess@wam.umd.edu}$^{a, \, b}$}}
\\[6mm]
\emph{
\centering
$^a$Center for String and Particle Theory-Dept.\ of Physics,
University of Maryland, \\[-2pt]
4150 Campus Dr., College Park, MD 20472,  USA
\\[8pt] 
and
\\[8pt] 
$^{b}$Department of Physics, Brown University,
\\[1pt]
Box 1843, 182 Hope Street, Barus \& Holley 545,
Providence, RI 02912, USA 
}
 \\*[90mm]
{ ABSTRACT}\\[4mm]
\parbox{142mm}{\parindent=2pc\indent\baselineskip=14pt plus1pt
We propose an explicit mathematical construction and plausibility arguments
for how spacetime chirality and Lorentz generators emerge for minimal, 
off-shell 4D, $\cal N$ = 1 supermultiplets by use of a 4.4.4.4 tessellation of 
Riemann surfaces based on plaquettes originating from
Coxeter Group BC${}_4$ adinkras.}
\end{center}
\vfill
\vfill
\clearpage

\section{Introduction}

Some times there occur difficulty in understanding how ``spacetime spin'' arises.  
An example of this is the, still not completely solved, problem of the ``proton spin 
crisis.''  Before 1987, it was expected the quarks were the principal carries of spin 
in the proton.  An experiment \cite{C1} that year revealed this was not supported by 
observation.  This raises the question, ``From where does the proton get its spin?''   
The spin of any relativistic spin-1/2 particle is closely related to the question of the 
Lorentz transformation properties of the particle.

At a recent meeting held at Brown University, a closely related question about 
adinkras was raised and addressed in an impromptu lecture by the author.  
Given the extemporaneous nature of this presentation, it was not a polished one 
and a main purpose of this note is to mitigate, remediate (any errors in the talk), and 
provide a formal set of definitions that seem needful in order to address the question, 
``From where do adinkras \cite{C2} get their spacetime spin?\footnote{This question 
was raised by J.\ Lukierski at a meeting in 2015 \cite{C2}.}''  As the talk was being 
given, it was also realized that many of the concepts introduced had not appeared 
previously in the literature.  So a secondary purpose of this work is to make these 
more widely available to any interested party.

\section{The Emergence of Lorentz Symmetry From $\bm {{\rm BC}}{}_4$ 
Adinkras: Stage I}

We begin with the adjoint representation of SO(4) and indicate
 its generators by $t_{{\rm I} {\rm J}}$, where this denotes the generator
whose infinitesimal effect is to rotate in the ${\rm I}-{\rm J}$ plane.  Let the
symbol $\L^{{\rm I}  {\rm J}}$ denotes six parameters so that $\L \cdot t$ is a
vector in the algebra of so(4).  Let ${{}^*} \L$ be the Hodge dual
of $\L$ when it is considered to be a 2-form.

The two quantities $\L ~\pm~ {}^* \L$ can now be shown to belong
to the two commuting su(2) algebras that occur in so(4).  The $\bm 
\alpha$ and $\bm \beta$ matrices in (\ref{AsBs})  represent the 
generators of these two distinct and commuting su(2) algebras.
The results in (\ref{Trcs}) and  (\ref{Trcs2}) are simply some of their properties.
We define
\be   \eqalign{
{\bm \a}{}^{\,\Hat 1}  ~&=~ \left[   { {{\bm \s}{}^2}}  \,  \otimes \, { {{\bm \s}{}^1}}  \right] ~\,~,~~ 
{\bm \a}{}^{\,\Hat 2} ~=~ \left[  {\bm {\rm I}}{}_2 \,  \otimes \, { {{\bm \s}{}^2}}  \right] ~~~~,~~ 
{\bm \a}{}^{\,\Hat 3} ~=~ \left[  {\bm \s}{}^2 \,  \otimes \, { {{\bm \s}{}^3}}  \right]  ~~, \cr
{\bm \b}{}^{\,\Hat 1}  ~&=~ \left[   { {{\bm \s}{}^1}} \,  \otimes \,   { {{\bm \s}{}^2}}  \right]  
\,~~, ~~ {\bm \b}{}^{\,\Hat 2}  ~=~  \left[     { {{\bm \s}{}^2}}  \,  \otimes \,  {\bm {\rm I}}{}_2
\right]  ~~~~, ~~ {\bm \b}{}^{\,\Hat 3} ~=~ \left[   { {{\bm \s}{}^3}}  \,  \otimes \,  { {{\bm \s}
{}^2}}   \right]  ~\,~,
} \label{AsBs}
\ee
where these matrices satisfy the identities
\be  \eqalign{ {~~~~~~~}
{\bm \a}^{\Hat {\rm I}} \, {\bm \a}^{\Hat {\rm K}} ~&=~ \delta{}^{{\Hat {\rm I}} \, {\Hat {\rm K}}} \, 
{\bm {\rm I}}{}_4 ~+~ i \, \epsilon{}^{{\Hat {\rm I}}  \, {\Hat {\rm K}} \, {\Hat {\rm L}}} \, {\bm 
\a}^{\Hat {\rm L}} ~~~, ~~~ 
{\bm \b}^{\Hat {\rm I}} \, {\bm \b}^{\Hat {\rm K}} ~=~ \delta{}^{{\Hat {\rm I}} \, {\Hat {\rm K}}} \,  
{\bm {\rm I}}{}_4 ~+~ i \, \epsilon{}^{{\Hat {\rm I}}  \, {\Hat {\rm K}} \, {\Hat {\rm L}}} \, 
{\bm \b}^{\Hat {\rm L}} ~~~,~~  
[\,  {\bm \a}^{\Hat {\rm I}}   ~,~ {\bm \b}^{\Hat {\rm J}}  \,  ] ~=~ 0 ~~,
 }   \label{Trcs}
\ee
\be  \eqalign{ {~~~~~~~}
& {~} {\rm {Tr}} \big( \,  {\bm \a}^{\Hat {\rm I}}  \, {\bm \a}^{\Hat {\rm J}}  \, \big) ~=~ {\rm {Tr}} 
\big( \, {\bm \b}^{\Hat {\rm I}}  \,  {\bm \b}^{\Hat {\rm J}} \, \big) ~=~ 4\,  \delta{}^{{\Hat {\rm I}} 
\, {\Hat {\rm J}}} ~~,~~ \ {\rm {Tr}} \big( \,  {\bm \a}^{\Hat {\rm I}}  \, {\bm \b}^{\Hat {\rm J}}  \, 
\big) ~=~ 0 ~~,   \cr
& {~~~~~~~~~~~~~~~~~~~~~} {\rm {Tr}} \big( \,  {\bm \a}^{\Hat {\rm I}}   \, \big) ~=~ {\rm 
{Tr}} \big( \, {\bm \b}^{\Hat {\rm I}}  \,  \, \big) ~=~ 0 ~~~.
}   \label{Trcs2}
\ee

Next we denote four traceless matrices by ${\bm {\Hat {\Sc K}}}{}^{\mu}$ (where $\mu$
= (0, 1, 2, 3)) and defined via the equations
\be \eqalign{ {~~~~~}
{\bm {\Hat {\Sc K}}}{}^{\mu}
&~=~ i \,   a^{\mu}{}^{\,\Hat {\rm I}} \, {\bm \a}{}^{\Hat {\rm I}}  ~+~ i \, b^{\mu}
{}^{\,\Hat {\rm I}} \, {\bm \b}{}^{\Hat {\rm I}}  ~+~  \, c^{\mu}{}^{{\Hat {\rm I}} \, 
{\Hat {\rm J}}} \,  {\bm \a}{}^{\Hat {\rm I}} \, {\bm \b}{}^{\Hat {\rm J}}
 ~~~,
}\label{Spn1}
\ee
where $a^{\mu}{}^{\,\Hat {\rm I}}$, $b^{\mu}{}^{\,\Hat {\rm I}}$, and $ c^{\mu}
{}^{{\Hat {\rm I}} \, {\Hat {\rm J}}}$ are 60 real constants.  To determine a solution 
space of values for these constants we impose the 
conditions
 \be \eqalign{ {~~~~~}
{\Big \{} \,   {\bm {\Hat {\Sc K}}}{}^{\mu} ~,~ {\bm {\Hat {\Sc K}}}{}^{\nu} \, {\Big \}  }
&~=~2  \, \eta{}^{\mu \, \nu} \, {\bm {\rm I}}{}_4 ~~~~,
}\label{Spn1aa}
\ee
where $\eta{}^{\mu \, \nu}$ is the Minkowski metric (our conventions are given
in the work \cite{G-1}}) for a space of one temporal and three spatial dimensions.  
As the matrices $ {\bm {\rm I}}{}_4 $, $i {\bm \a}^{\Hat {\rm I}}$, $i {\bm \b
}^{\Hat {\rm I}}$, and $ {\bm \a}^{\Hat {\rm I}} {\bm \b}^{\Hat {\rm J}}$ constitute a 
complete basis for constructing any real 4 $\times$4 matrix, this guarantees 
solutions must exist.

Upon calculating the left hand side, we find
 \be \eqalign{ {\,}
{\Big \{} \,   {\bm {\Hat {\Sc K}}}{}^{\mu} ~,~ {\bm {\Hat {\Sc K}}}{}^{\nu} \, {\Big \}  }
&~=~ - \,  a^{\mu}{}^{\,\Hat {\rm I}} \,  \,  a^{\nu}{}^{\,\Hat {\rm J}} \,
{\Big \{} \,   {\bm \a}{}^{\Hat {\rm I}} ~,~ {\bm \a}{}^{\Hat {\rm J}}  \, {\Big \}  } - \, 
  b^{\mu}{}^{\,\Hat {\rm I}} \,  \,  b^{\nu}{}^{\,\Hat {\rm J}} \, {\Big \{}
 \,   {\bm \b}{}^{\Hat {\rm I}} ~,~ {\bm \b}{}^{\Hat {\rm J}}  \, {\Big \}  }  \cr
&~~~~~~+~ i \,  {\big (}\,   b^{\mu}{}^{\,\Hat {\rm I}} \,  c^{\nu}
{}^{\,\Hat {\rm K} \,\Hat {\rm L}} ~+~ b^{\nu}{}^{\,\Hat {\rm I}} \,  c^{\mu}{}^{\,
\Hat {\rm K} \,\Hat {\rm L}} \, {\big )} \, {\Big \{} \,   {\bm \b}{}^{\Hat {\rm I}} 
~,~ {\bm \a}{}^{\Hat {\rm K}} \, {\bm \b}{}^{\Hat {\rm L}} \, {\Big \}  }   \cr 
&~~~~~~+~ i \,  {\big (}\,   a^{\mu}{}^{\,\Hat {\rm I}} \,  c^{\nu}
{}^{\,\Hat {\rm K} \,\Hat {\rm L}} ~+~ a^{\nu}{}^{\,\Hat {\rm I}} \,  c^{\mu}{}^{\,
\Hat {\rm K} \,\Hat {\rm L}} \, {\big )} \, {\Big \{} \,   {\bm \a}{}^{\Hat {\rm I}} 
~,~ {\bm \a}{}^{\Hat {\rm K}} \, {\bm \b}{}^{\Hat {\rm L}} \, {\Big \}  }  \cr
&~~~~~~-~  {\big (}\, a^{\mu}{}^{\,\Hat {\rm I}} \,  b^{\nu}{}^{\,\Hat 
{\rm J}} ~+~ a^{\nu}{}^{\,\Hat {\rm I}} \,  \,  b^{\mu}{}^{\,\Hat {\rm J}}  {\big )} \,
{\Big \{} \,   {\bm \a}{}^{\Hat {\rm I}} ~,~ {\bm \b}{}^{\Hat {\rm J}}  \, {\Big \}  }\cr
&~~~~~~+~  \,  c^{\mu}{}^{\,\Hat {\rm I} \,\Hat {\rm J}} \,  c^{\nu}
{}^{\,\Hat {\rm K} \,\Hat {\rm L}} \, {\Big \{} \,   {\bm \a}{}^{\Hat {\rm I}} \,  {\bm \b}
{}^{\Hat {\rm J}}  ~,~ {\bm \a}{}^{\Hat {\rm K}} \, {\bm \b}{}^{\Hat {\rm L}} \, {\Big \}  }
 ~~~,
}\label{Spn1a}
\ee

\be \eqalign{ {~~~~~}
{\Big \{} \,   {\bm {\Hat {\Sc K}}}{}^{\mu} ~,~ {\bm {\Hat {\Sc K}}}{}^{\nu} 
\, {\Big \}  } &~=~ - \, 2 \, {\big (} \,  a^{\mu}{}^{\,\Hat {\rm I}} \, 
a^{\nu}{}^{\,\Hat {\rm I}} ~+~ b^{\mu}{}^{\,\Hat {\rm I}} \,  \,  b^{\nu}{}^{\,
\Hat {\rm I}} \, {\big )}   \,  {\bm {\rm I}}{}_4    \cr
&~~~~~~+~ i \, 2 {\big (}\,   b^{\mu}{}^{\,\Hat {\rm I}} \,  c^{\nu}
{}^{\,\Hat {\rm J} \,\Hat {\rm I}} ~+~ b^{\nu}{}^{\,\Hat {\rm I}} \,  c^{\mu}{}^{\,
\Hat {\rm J} \,\Hat {\rm I}} \, {\big )} \, {\bm \a}{}^{\Hat {\rm J}}  \cr
&~~~~~~+~ i \, 2 {\big (}\,   a^{\mu}{}^{\,\Hat {\rm I}} \,  c^{\nu}
{}^{\,\Hat {\rm I} \,\Hat {\rm J}} ~+~ a^{\nu}{}^{\,\Hat {\rm I}} \,  c^{\mu}{}^{\,
\Hat {\rm I} \,\Hat {\rm J}} \, {\big )} \,   {\bm \b}{}^{\Hat {\rm J}} \cr
&~~~~~~-~ 2 \, {\big (}\, a^{\mu}{}^{\,\Hat {\rm I}} \,  b^{\nu}{}^{\,\Hat 
{\rm J}} ~+~ a^{\nu}{}^{\,\Hat {\rm I}} \,  \,  b^{\mu}{}^{\,\Hat {\rm J}}  {\big )} \,
{\bm \a}{}^{\Hat {\rm I}} \, {\bm \b}{}^{\Hat {\rm J}}  \cr
&~~~~~~+~  c^{\mu}{}^{\,\Hat {\rm I} \,\Hat {\rm J}} \, c^{\nu}
{}^{\,\Hat {\rm K} \,\Hat {\rm L}} \, {\big (} \,   {\bm \a}{}^{\Hat {\rm I}} \,  {\bm 
 \a}{}^{\Hat {\rm K}} \,{\bm \b}{}^{\Hat {\rm J}} \, {\bm \b}{}^{\Hat {\rm L}}  ~+~
{\bm \a}{}^{\Hat {\rm K}} \, {\bm \a}{}^{\Hat {\rm I}}  \, {\bm \b}{}^{\Hat {\rm L}}
\, {\bm \b} {}^{\Hat {\rm J}} \, {\big )  }
  ~~~,
}\label{Spn1b}
\ee

 \be \eqalign{ {~}
{\Big \{} \,   {\bm {\Hat {\Sc K}}}{}^{\mu} ~,~ {\bm {\Hat {\Sc K}}}{}^{\nu} 
\, {\Big \}  } &~=~ - \, 2 \, {\big (} \,  a^{\mu}{}^{\,\Hat {\rm I}} \, 
a^{\nu}{}^{\,\Hat {\rm I}} ~+~ b^{\mu}{}^{\,\Hat {\rm I}} \,  \,  b^{\nu}{}^{\,
\Hat {\rm I}}  ~-~  c^{\mu}
{}^{\,\Hat {\rm I} \,\Hat {\rm J}} \, c^{\nu}{}^{\,\Hat {\rm I} \,\Hat {\rm J}} \,  
\, {\big )}   \,  {\bm {\rm I}}{}_4   \cr
&~~~~~~+~ i \, 2 \, {\big (}\,   b^{\mu}{}^{\,\Hat {\rm I}} \,  c^{\nu}
{}^{\,\Hat {\rm J} \,\Hat {\rm I}} ~+~ b^{\nu}{}^{\,\Hat {\rm I}} \,  c^{\mu}{}^{\,
\Hat {\rm J} \,\Hat {\rm I}} \, {\big )} \, {\bm \a}{}^{\Hat {\rm J}}    \cr
&~~~~~~+~ i \, 2  \, {\big (}\,   a^{\mu}{}^{\,\Hat {\rm I}} \,  c^{\nu}
{}^{\,\Hat {\rm I} \,\Hat {\rm J}} ~+~ a^{\nu}{}^{\,\Hat {\rm I}} \,  c^{\mu}{}^{\,
\Hat {\rm I} \,\Hat {\rm J}} \, {\big )} \,   {\bm \b}{}^{\Hat {\rm J}} \cr
&~~~~~~-~  2 \, {\big (}\, a^{\mu}{}^{\,\Hat {\rm I}} \,  b^{\nu}{}^{\,\Hat 
{\rm J}} ~+~ a^{\nu}{}^{\,\Hat {\rm I}} \,  \,  b^{\mu}{}^{\,\Hat {\rm J}}  {\big )} \,
{\bm \a}{}^{\Hat {\rm I}} \, {\bm \b}{}^{\Hat {\rm J}} \cr
&~~~~~~-~  2 \, c^{\mu}{}^{\,\Hat {\rm I} \,\Hat {\rm J}} \, c^{\nu}{}^{\,\Hat {\rm 
K} \,\Hat {\rm L}} \,  \e{}^{\Hat {\rm I}}{}^{\Hat {\rm K}}{}^{\Hat {\rm R}} \, \e{}^{
 \Hat {\rm J}}{}^{\Hat {\rm L}}{}^{\Hat {\rm S}} \,  {\bm \a}{}^{\Hat {\rm R}} 
\, {\bm \b}{}^{\Hat {\rm S}}  
  ~~~.  {~~~~~~~~~~~~~~\,~}
}\label{Spn1c}
\ee

The condition in (\ref{Spn1aa}) is thus seen to be equivalent to 160
quadratic polynomial constraints that define an algebraic variety whose 
explicit form is given by,
\be  \eqalign{  {~~~~~}
\eta{}^{\mu \, \nu} ~&=~  - \,  {\big (} \,  a^{\mu}{}^{\,\Hat {\rm I}} \, 
a^{\nu}{}^{\,\Hat {\rm I}} ~+~ b^{\mu}{}^{\,\Hat {\rm I}} \,  \,  b^{\nu}{}^{\, \Hat 
{\rm I}} \, {\big )} ~+~ c^{\mu}{}^{\,\Hat {\rm I} \,\Hat {\rm J}} \, c^{\nu}
{}^{\,\Hat {\rm I} \,\Hat {\rm J}} ~~~~,  ~~~~~~~~~~~~~~~~(\# ~{\rm{of}}~
{{\rm {constraints}:}~ 10})  \cr
0 ~&=~   a^{\mu}{}^{\,\Hat {\rm I}} \, c^{\nu}{}^{\,\Hat {\rm I} \,\Hat {\rm J}} 
~+~ a^{\nu}{}^{\,\Hat {\rm I}} \,  c^{\mu}{}^{\,\Hat {\rm I} \,\Hat {\rm J}} ~~\,~,~  
~~~~~~~~~~~~~~~~~~~~~~~~~~~~~~~~~~~\,~~~(\# ~{\rm{of}}~{{\rm {constraints}}: 
30})  \cr
0 ~&=~ b^{\mu}{}^{\,\Hat {\rm I}} \, c^{\nu}{}^{\,\Hat {\rm J} \,\Hat {\rm I}} ~+~ 
b^{\nu}{}^{\,\Hat {\rm I}} \,  c^{\mu}{}^{\, \Hat {\rm J} \,\Hat {\rm I}}  ~~~~,  
~~~~~~~~~~~~~~~~~~~~~~~~~~~~~~~\,~~~~\,~~~(\# ~{\rm{of}}~{{\rm {constraints}}: ~30})
 \cr
0 ~&=~  c^{\mu}{}^{\,\Hat {\rm I} \,\Hat {\rm J}} \, c^{\nu}{}^{\,\Hat {\rm K} \,\Hat {\rm L}} 
\,  \e{}^{\Hat {\rm I}}{}^{\Hat {\rm K}}{}^{\Hat {\rm R}} \, \e{}^{\Hat {\rm J}}{}^{\Hat {\rm 
L}}{}^{\Hat {\rm S}} ~+~ {\big (}\, a^{\mu}{}^{\,\Hat {\rm R}} \,  b^{\nu}{}^{\,\Hat {\rm S}} 
~+~ a^{\nu}{}^{\,\Hat {\rm R}} \,  \,  b^{\mu}{}^{\,\Hat {\rm S}}  {\big )} ~~~~,  
\cr
~&=~   \d{}^{\Hat {\rm R}}{}^{\Hat {\rm S}} \, ( c^{\mu}{}^{\,\Hat {\rm I} \,\Hat {\rm I}} 
\, c^{\nu}{}^{\,\Hat {\rm J} \,\Hat {\rm J}} \,-\, c^{\mu}{}^{\,\Hat {\rm I} \,\Hat {\rm J}} \, 
c^{\nu}{}^{\,\Hat {\rm J} \,\Hat {\rm I}}   )  \cr  
&~~~~+~ (  c^{\mu}{}^{\,\Hat {\rm S} \,\Hat {\rm J}} \, c^{\nu}{}^{\,\Hat {\rm J} \,\Hat {\rm 
R}}  \,-\,  c^{\mu}{}^{\,\Hat {\rm I} \,\Hat {\rm I}} \, c^{\nu}{}^{\,\Hat {\rm R} \,\Hat {\rm S}} 
 )  \cr 
&~~~~+~ (  c^{\nu}{}^{\,\Hat {\rm R} \,\Hat {\rm J}} \, c^{\mu}{}^{\,\Hat {\rm J} \,\Hat {\rm 
S}}  \,-\,  c^{\nu}{}^{\,\Hat {\rm I} \,\Hat {\rm I}} \, c^{\mu}{}^{\,\Hat {\rm S} \,\Hat {\rm R}} 
 )  \cr 
&~~~~+~ {\big (}\, a^{\mu}{}^{\,\Hat {\rm R}} \,  b^{\nu}{}^{\,\Hat {\rm S}} 
~+~ a^{\nu}{}^{\,\Hat {\rm R}} \,  \,  b^{\mu}{}^{\,\Hat {\rm S}}  {\big )} ~~~~,
~~~~~~~~~~~~~~~~~~~~~~~~~~~~~~~~(\# ~{\rm{of}}~{{\rm {constraints}}: ~90}) 
}  \label{CNstr} \ee
to impose constraints on the 60 constants represented as $a^{\nu}{}^{\,\Hat {\rm I}}$,
$b^{\nu}{}^{\,\Hat {\rm I}}$, and $c^{\nu}{}^{\,\Hat {\rm J} \,\Hat {\rm I}} $.  

So the next question becomes what structures ``native and natural'' to
adinkras associated with BC${}_4$ can give rise to the basis seen in
(\ref{Spn1}) above?

\section{The Emergence of Lorentz Symmetry From $\bm {{\rm BC}}{}_4$ Adinkras: Stage II}

To every 4-four color, 4-open node, 4-closed node adinkra, such as shown in the figures below,

$$
\vCent
{\setlength{\unitlength}{1mm}
\begin{picture}(-20,0)
\put(-46,-22){\includegraphics[width=3.4in]{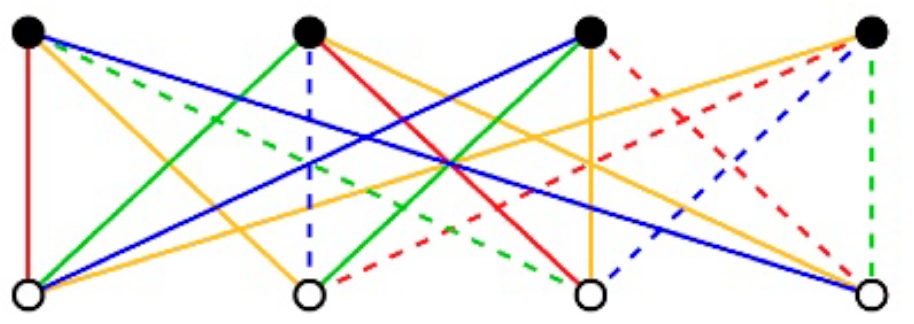}}
\end{picture}}
\nonumber
$$
\vskip.6in
$ {~~~~~~~~~~~~~~~~~~~~~~~~~~~~~~~~~~~~~~~~~~~~} {\bm {\rm {Figure ~ 1:}}} 
\bm {~~(\cal R )} ~=~  \bm {(CM)} $

$$
\vCent
{\setlength{\unitlength}{1mm}
\begin{picture}(-20,0)
\put(-46,-24){\includegraphics[width=3.4in]{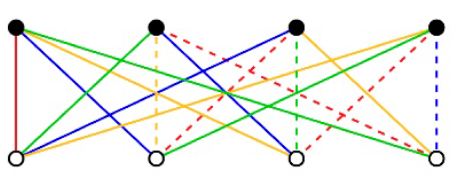}}
\end{picture}}
\nonumber
$$
\vskip.6in
$ {~~~~~~~~~~~~~~~~~~~~~~~~~~~~~~~~~~~~~~~~~~~~} {\bm {\rm {Figure ~ 2:}}} 
\bm {~~(\cal R )} ~=~  \bm {(TM)} $

$$
\vCent
{\setlength{\unitlength}{1mm}
\begin{picture}(-20,0)
\put(-46,-22){\includegraphics[width=3.4in]{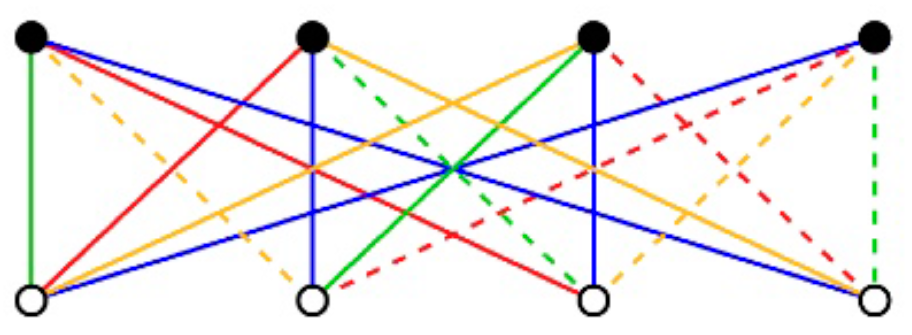}}
\end{picture}}
\nonumber
$$
\vskip.6in
$ {~~~~~~~~~~~~~~~~~~~~~~~~~~~~~~~~~~~~~~~~~~~~} {\bm {\rm {Figure ~ 3:}}} 
\bm {~~(\cal R )} ~=~  \bm {(VM)} $ 
\vskip.2in
\noindent
there corresponds a set of ``L-matrices'' and ``R-matrices''  ${\bm {\rm L}}{}^{(\cal R)}_\rI\,$ 
and ${\bm {\rm R}}{}^{(\cal R)}_\rI$ that satisfy the ``Garden Algebra.''
\be { \eqalign{
{\bm {\rm L}}{}^{(\cal R)}_\rI  \, {\bm {\rm R}}{}^{(\cal R)}_\rJ ~+~ {\bm {\rm L
 }}{}^{(\cal R)}_\rJ \, {\bm {\rm R}}{}^{(\cal R)}_\rI  ~&=~ 2\,\d_{\rI\rJ}\,
{\bm {\rm I}}{}_{d \times d} ~~,\cr
{\bm {\rm  R}}{}^{(\cal R)}_\rI   \,  {\bm {\rm L}}{}^{(\cal R)}_\rJ ~+~
{\bm {\rm R}}{}^{(\cal R)}_\rJ   \,  {\bm {\rm L}}{}^{(\cal R)}_\rI    ~&=~ 2\,\d_{\rI\rJ}
 {\bm {\rm I}}{}_{d \times d} ~~,  \cr
~~~ {\bm {\rm  R}}{}^{(\cal R)}_\rI  ~=~ [ \,  {\bm {\rm L}}{}^{(\cal R)}_\rI \, ]{}^{-1}
&~~.
}}\label{GarDNAlg2}
\ee
and the fermionic holoraumy matrix ${\bm {\Tilde V}}$ ${}_{\rI\rJ}^{(\cal R)}$ for each 
representation is defined via the equation
\be { \eqalign{
{\bm {\rm  R}}{}^{(\cal R)}_\rI   \,  {\bm {\rm L}}{}^{(\cal R)}_\rJ ~-~
{\bm {\rm R}}{}^{(\cal R)}_\rJ   \,  {\bm {\rm L}}{}^{(\cal R)}_\rI   
&= i \, 2\,  {\bm {\Tilde V}}{}_{\rI\rJ}^{(\cal R)} ~~.  }}\label{GarDVs}
\ee
Due to the definitions in (\ref{GarDNAlg2}), it follows the matrices ${\bm {\Tilde V}} {}_{\rI\rJ
}^{(\cal R)}$ are elements in the so(4) algebra and thus for any two representation $({\cal 
R})$ and $({\cal R^{\prime}})$ we may write
 \be   \eqalign{ {~~~~~~~}
{\bm {\Tilde V}{}_{\rI\rJ}^{(\cal R)}} ~=~ &\ell^{({\cal R}){\Hat {\rm I}}}_{\rI\rJ}\, {\bm {\a}}
{}^{\,\Hat {\rm I}}    ~+~   {{\Tilde \ell}^{(\cal R)}}_{\rI\rJ}{}^{\Hat {\rm I}}\, {\bm {\b}}{}^{\,
\Hat {\rm I}}    ~~~,~~~  {\bm {\Tilde V}{}_{\rI\rJ}^{({\cal R}^{\prime})}} ~=~ \ell^{({{\cal 
R}^{\prime}}){\Hat {\rm I}}}_{\rI\rJ}\, {\bm {\a}}{}^{\,\Hat {\rm I}} ~+~   {{\Tilde \ell}^{({\cal R
}^{\prime})}}_{\rI\rJ}{}^{\Hat {\rm I}}\,  {\bm {\b}}{}^{\,\Hat {\rm I}}    ~~~, }  \label{Veqz2}
\ee
for some set of coefficients $\ell^{({\cal R}){\Hat {\rm I}}}_{\rI\rJ}$, ${{\Tilde \ell}^{(\cal 
R)}}_{\rI\rJ}{}^{\Hat {\rm I}}$,  $\ell^{({\cal R}^{\prime})){\Hat {\rm I}}}_{\rI\rJ}$, and ${{\Tilde 
\ell}^{({\cal R}^{\prime})}}_{\rI\rJ}{}^{\Hat {\rm I}}$.

Next, we consider two adinkra representations $({\cal R})$ and $({\cal R^{\prime}})$ associated
with BC${}_4$ and construct a matrix ${\bm k}{}^{\mu}\langle \, ({\cal R}) {\big |} ({\cal R}^{\prime
}) \, \rangle $ via the equation
\be {~~~~~~~~~}
{\bm k}{}^{\mu} \langle \, ({\cal R}) {\big |} ({\cal R}^{\prime}) \, \rangle ~=~  i \, A^{\mu}{}^{\,
{\rm I}}{}^{\,   {\rm J}}  \, {\bm { \Tilde{V}}}{}_{{\, {\rm I}}{\, {\rm J}} }^{({\cal R})}  ~+~  i 
\, B^{\mu}{}^{\, {\rm I}}{}^{\,   {\rm J}}  \, {\bm { \Tilde{V}}}{}_{{\,   {\rm I}}{\,   {\rm J}} }^{
({\cal R}^{\prime})} ~+~ C^{\mu}{}^{\,   {\rm I}}{}^{\,   {\rm J}}{}^{\,   {\rm K}}{}^{\,   {\rm 
L}} \, {\bm { \Tilde{V}}}{}_{{\,   {\rm I}}{\,   {\rm J}} }^{({\cal R})} \, {\bm { \Tilde{V}}}
{}_{{\,   {\rm K}}{\,   {\rm L}} }^{({\cal R}^{\prime})}
  ~~~, ~~~~~~~~
   \label{Spn1ad0}
\ee
which is seen to be a quadratic polynomial in the fermionic holoraumy matrices.
By use of the equations in (\ref{Veqz2}), ${\bm k}{}^{\mu}\langle \, ({\cal R}) {\big |} 
({\cal R}^{\prime}) \, \rangle $
becomes
\be    \eqalign{ {~~~~~~}
{\bm k}{}^{\mu} \langle \, ({\cal R}) {\big |} ({\cal R}^{\prime}) \, \rangle
~=~  i \, &{\Big [} \,
 A^{\mu}{}^{\,   {\rm I}}{}^{\,   {\rm J}}  {\Big (} \, \ell^{({\cal R}){\Hat {\rm I}}}_{\rI\rJ} 
~+~ \ell^{({\cal R}^{\prime}) {\Hat {\rm I}}}_{\rI\rJ} {\Big )} ~+~ C^{\mu}{}^{\,{\rm I}}
{}^{\,   {\rm J}}{}^{\,   {\rm K}}{}^{\,   {\rm L}} \ell^{({\cal R}){\Hat {\rm K}}}_{\rI\rJ} \, 
\ell^{({\cal R}^{\prime}) {\Hat {\rm L}}}_{\rK\rL}\,  \e{}^{\Hat {\rm K}}{}^{\Hat {\rm L}}{
}^{\Hat {\rm I}} \, {\Big ]}  \, {\bm {\a}}{}^{\,\Hat {\rm I}}  ~+~  \cr  
 i \, &{\Big [}  \, 
  B^{\mu}{}^{\,   {\rm I}}{}^{\,   {\rm J}}  {\Big (} \, {\Tilde \ell}^{({\cal R}){\Hat 
{\rm I}}}_{\rI\rJ} ~+~ {\Tilde \ell}^{({\cal R}^{\prime}) {\Hat {\rm I}}}_{\rI\rJ} {\Big )} ~+~ C^{\mu}
{}^{\,   {\rm I}}{}^{\,   {\rm J}}{}^{\,   {\rm K}}{}^{\,   {\rm L}} {\Tilde \ell}^{({\cal R}){\Hat {\rm K}}
}_{\rI\rJ} \, {\Tilde \ell}^{({\cal R}^{\prime}) {\Hat {\rm L}}}_{\rK\rL}\,  \e{}^{\Hat {\rm K}}{}^{\Hat 
{\rm L}}{}^{\Hat {\rm I}}   {\Big ]}  \, {\bm {\b}}{}^{\,\Hat {\rm I}}  \, ~+~  \cr
&~~C^{\mu}{}^{\,   {\rm I}}{}^{\,   {\rm J}}{}^{\,   {\rm K}}{}^{\,   {\rm L}}  {\Big [} \,
\ell^{({\cal R}){\Hat {\rm I}}}_{\rI\rJ} \,  {\Tilde \ell}^{({\cal R}^{\prime}) {\Hat 
{\rm J}}}_{\rK\rL} ~+~
{\Tilde \ell}^{({\cal R}){\Hat {\rm J}}}_{\rI\rJ} \,  {\ell}^{({\cal R}^{\prime}) {\Hat {\rm I}}}_{\rK\rL}
  {\Big ]}  \, {\bm {\a}}{}^{\,\Hat {\rm I}} 
  \, {\bm {\b}}{}^{\,\Hat {\rm J}} 
  ~~~. ~~
  } \label{Spn1ad}
\ee
This has exactly the form of (\ref{Spn1}) where the $a$, $b$, and $c$ coefficients there 
are related to the the $\ell$, and $\Tilde \ell$ parameters in (\ref{Veqz2}) and $A$, $B$, 
and $C$ coefficients in (\ref{Spn1ad0}) as well through the relationships
\be    \eqalign{ {~~~~~~~~~~~~~}
a^{\mu}{}^{\,\Hat {\rm I}}   ~=~   &{\Big [} \,
 A^{\mu}{}^{\,   {\rm I}}{}^{\,   {\rm J}}  {\Big (} \, \ell^{({\cal R}){\Hat {\rm I}}}_{\rI\rJ} 
~+~ \ell^{({\cal R}^{\prime}) {\Hat {\rm I}}}_{\rI\rJ} {\Big )} ~+~ C^{\mu}{}^{\,{\rm I}}
{}^{\,   {\rm J}}{}^{\,   {\rm K}}{}^{\,   {\rm L}} \ell^{({\cal R}){\Hat {\rm K}}}_{\rI\rJ} \, 
\ell^{({\cal R}^{\prime}) {\Hat {\rm L}}}_{\rK\rL}\,  \e{}^{\Hat {\rm K}}{}^{\Hat {\rm L}}{
}^{\Hat {\rm I}} \, {\Big ]}    ~~~,  \cr  
b^{\mu}{}^{\,\Hat {\rm I}}   ~=~ &{\Big [}  \,  B^{\mu}{}^{\,   {\rm I}}{}^{\,   {\rm J}}  {\Big (} \, 
{\Tilde \ell}^{({\cal R}){\Hat {\rm I}}}_{\rI\rJ} ~+~ {\Tilde \ell}^{({\cal R}^{\prime}) {\Hat {
\rm I}}}_{\rI\rJ} {\Big )} ~+~ C^{\mu}{}^{\,   {\rm I}}{}^{\,   {\rm J}}{}^{\,   {\rm K}}{}^{\, {\rm 
L}} {\Tilde \ell}^{({\cal R}){\Hat {\rm K}}}_{\rI\rJ} \, {\Tilde \ell}^{({\cal R}^{\prime}) {\Hat 
{\rm L}}}_{\rK\rL}\,  \e{}^{\Hat {\rm K}}{}^{\Hat {\rm L}}{}^{\Hat {\rm I}}   {\Big ]}    \, ~~~,  \cr
c^{\mu}{}^{{\Hat {\rm I}} \, {\Hat {\rm J}}} ~=~ &~C^{\mu}{}^{\,   {\rm I}}{}^{\,   {\rm J}}{}^{\,  
{\rm K}}{}^{\,   {\rm L}}  {\Big [}\, \,\ell^{({\cal R}){\Hat {\rm I}}}_{\rI\rJ} \,  {\Tilde \ell}^{({\cal R}^{
\prime}) {\Hat {\rm J}}}_{\rK\rL} ~+~ {\Tilde \ell}^{({\cal R}){\Hat {\rm J}}}_{\rI\rJ} \,  {\ell}^{({\cal 
R}^{\prime}) {\Hat {\rm I}}}_{\rK\rL} \, \, {\Big ]} ~\equiv~ C^{\mu}{}^{\, {\rm I}}{}^{\, {\rm J}}{}^{\,  
{\rm K}}{}^{\,   {\rm L}} \, \D{}_{\rI\rJ\rK\rL}^{\Hat {\rm I} \, \Hat {\rm J}}
\langle \, ({\cal R}) {\big |} ({\cal R}^{\prime}) \, \rangle
  ~~~. ~~
  } \label{Spn1ad2}
\ee
Given the equations in (\ref{Spn1ad2}), we can now formulate a conjecture about the 
construction of Dirac Gamma matrices on the basis of adinkras related to BC${}_4$.

$~~~~$ {\it{Conjecture One}}: 
\newline $~~~~~~~~~~$ {\it{Let}} $({\cal R})$ and $({\cal R}^{\prime})$ {\it{denote any 
four color adinkra graphs associated with}} BC${}_4$.   {\it{To each}} \newline 
$~~~~~~~~~~$ {\it{such graph, there exist six associated ``fermionic holoraumy 
matrices''}} ${\bm {\Tilde V}{}_{\rI\rJ}^{({\cal R})}}$ and ${\bm {\Tilde V}{}_{\rI\rJ}^{({\cal 
R}^{\prime})}}$.  \newline $~~~~~~~~~~$ {\it{If}}  $ \D{}_{\rI\rJ\rK\rL}^{\Hat {\rm I} \, \Hat 
{\rm J}} \langle ({\cal R}) {\big |} ({\cal R}^{\prime}) \rangle$ = $0$,  {\it{then the equation}}
\be  \eqalign{ {~~~~~~~~~~~~~}
{\bm k}{}^{\mu} \langle \, ({\cal R}) {\big |} ({\cal R}^{\prime}) \, \rangle \, 
{\bm k}{}^{\nu} \langle \, ({\cal R}) {\big |} ({\cal R}^{\prime}) \, \rangle ~+~
{\bm k}{}^{\nu} \langle \, ({\cal R}) {\big |} ({\cal R}^{\prime}) \, \rangle \,
{\bm k}{}^{\mu} \langle \, ({\cal R}) {\big |} ({\cal R}^{\prime}) \, \rangle  ~=~
2 \, \eta{}^{\mu}{}^{\nu} \,  \, {\bm {\rm I}}{}_4 ~~~,
} \label{C1} \ee
$~~~~~~~~~$ {\it{possesses no solutions.}} \vskip4pt
\noindent
An argument for the validity of this is suggested by the following observations.

We can look back to the first equation in (\ref{CNstr}).  This equation makes it
clear for the diagonal entries of the Minkowski metric on the left-hand side
to be indefinite requires that the $c{}^{\m}{}^{\Hat I}{}^{\Hat J}$ be non-vanishing.
Otherwise the diagonal entries of $\eta{}^{\m}{}^{\nu}$ must be negative definite.
In order to have $c{}^{\m}{}^{\Hat I}{}^{\Hat J}$ be non-vanishing, it must be the
case that $ \D{}_{\rI\rJ\rK\rL}^{\Hat {\rm I} \, \Hat {\rm J}} \langle ({\cal R}) {\big |} 
({\cal R}^{\prime}) \rangle$ must be non-vanishing.

One can calculate the quantity $ \D{}_{\rI\rJ\rK\rL}^{\Hat {\rm I} \, \Hat {\rm J}}
\langle \, ({\cal R}) {\big |} ({\cal R}^{\prime}) \, \rangle$ on the $(CM)$, $(TM)$ and $(VM)$ 
pairs, to find when it is vanishes.  When it vanishes, then $c^{\mu}{}^{{\Hat {\rm I}} \, 
{\Hat {\rm J}}}$ vanishes, by looking at the purely diagonal entries in the first
equation of (\ref{CNstr}), one sees the impossibility to satisfy this equation due to 
the indefinite nature of the Minkowski metric.  This brings us to a second
conjecture.

$~~~~$ {\it{Conjecture Two}}: 
\newline $~~~~~~~~~~$ {\it{Let}} $({\cal R})$ and $({\cal R}^{\prime})$ {\it{denote any 
four color adinkra graphs associated with}} BC${}_4$.   {\it{If the}} \newline 
$~~~~~~~~~~$ {\it{condition}} $\D{}_{\rI\rJ\rK\rL}^{\Hat {\rm I} \, \Hat {\rm J}} \langle \, 
({\cal R}) {\big |} ({\cal R}^{\prime}) \, \rangle$ $\ne$ $0$,  {\it{is satisfied, then the equation 
in}} (\ref{Spn1aa}) {\it{possesses}}
 \newline 
$~~~~~~~~~~$ {\it{multiple solutions.}} \vskip4pt
\noindent
An argument for the validity of this conjecture is seen by simply carrying out the
explicit calculations indicated to examine  $\D{}_{\rI\rJ\rK\rL}^{\Hat {\rm I} \, \Hat 
{\rm J}} \langle \, ({\cal R}) {\big |} ({\cal R}^{\prime}) \, \rangle$ among the adinkras in 
Fig.\ \#1, Fig.\ \#2, and Fig.\ \#3.  This will be done shortly.  We emphasize these 
are conjectures as we expect these to hold over all the adinkras associated with 
BC${}_4$.  The work in \cite{adnkBillions} shows this number is 36,864 adinkras.  

As Boolean Factors $\times$ permutations cycles (BFPC) \footnote{
Here we use the ``read down'' convention for relating cycle notation 
to matrix notation of permutations as discussed in the work of \cite{adnkBillions}.} 
the three adinkras correspond to,
\be  \eqalign{  {\rm {Red}}~~~&~~~~~~~~~~~~~~~~~~~~~~~{\rm {Green}}
~~~~~~~~~~~~~~~~~~~~~~~{\rm {Blue}}
~~~~~~~~~~~~~~~~~~~~~~~{\rm {Yellow}} \cr
{\bm {\rm L}}{}^{(CM)}_{1}   ~&=~ (10)_b (234) ~\,~,~ {\bm {\rm L}}{}^{(CM)}_{2}
~=~ (12)_b (132) ~,~ {\bm {\rm L}}{}^{(CM)}_{3} ~=~ (6)_b (143)
\,~,~~ {\bm {\rm L}}{}^{(CM)}_{4} ~=~ (0)_b (124)
~~~,   \cr
{\bm {\rm L}}{}^{(TM)}_{1}   ~&=~  (14)_b (243)  ~\,~,~ {\bm {\rm L}}{}^{(TM)}_{2}
~=~  (4)_b (142)  \,~~,~ {\bm {\rm L}}{}^{(TM)}_{3} ~=~  (8)_b (123) 
~~,~~ {\bm {\rm L}}{}^{(TM)}_{4} ~=~  (2)_b (134) 
~~~,  \cr
{\bm {\rm L}}{}^{(VM)}_{1}   ~&=~  (10)_b (1342) ~,~ {\bm {\rm L}}{}^{(VM)}_{2}
~=~  (12)_b (23)  ~\,~,~ {\bm {\rm L}}{}^{(VM)}_{3} ~=~  (0)_b (14) 
~~~,~~ {\bm {\rm L}}{}^{(VM)}_{4} ~=~  (6)_b (1243) 
\,~, }
\label{BFPVM}
\ee
that were introduced in the work of \cite{permutadnk}.  Above, each column
of L-matrices is listed vertically under the rainbow color shown in the corresponding
adinkra. In more explicit form these become  \newline
\noindent
$\bm {CM~Adinkra~L-Matrices}$
$$  {
{\bm {\rm L}}{}^{(CM)}_{1}   ~=~
\left[\begin{array}{cccc}
~1 & ~~0 &  ~~0  &  ~~0 \\
~0 & ~~0 &  ~~0  &  ~-\, 1 \\
~0 & ~~1 &  ~~0  &  ~~0 \\
~0 & ~~0 &  ~-\, 1  &  ~~0 \\
\end{array}\right] ~~~,~~~
{\bm {\rm L}}{}^{(CM)}_{2}   ~=~
\left[\begin{array}{cccc}
~0 & ~~1 &  ~~0  &  ~ \, \, 0 \\
~0 & ~~ 0 &  ~~1  &  ~~0 \\
-\, 1 & ~~ 0 &  ~~0  &  ~~0 \\
~ 0 & ~~~0 &  ~~0  &   -\, 1 \\
\end{array}\right]  ~~~, }
$$
\be  {
  {\bm {\rm L}}{}^{(CM)}_{3}   ~=~
\left[\begin{array}{cccc}
~0 & ~~0 &  ~~1  &  ~~0 \\
~0 & ~- \, 1 &  ~~0  &  ~~0 \\
~0 & ~~0 &  ~~0  &  -\, 1 \\
~1 & ~~0 &  ~~0  &  ~~0 \\
\end{array}\right] ~~~,~~~
  {\bm {\rm L}}{}^{(CM)}_{4}   ~=~
\left[\begin{array}{cccc}
~0 & ~~0 &  ~~0  &  ~ \, \, 1 \\
~1 & ~~ 0 &  ~~0  &  ~~0 \\
~0 & ~~ 0 &  ~~1  &  ~~0 \\
~ 0 & ~~~1 &  ~~0  &   ~~0  \\
\end{array}\right]  ~~~~,  }
 \label{chiDoF}
\ee 
$\bm {TM~Adinkra~L-Matrices}$
$$  { 
  {\bm {\rm L}}{}^{(TM)}_{1}   ~=~
\left[\begin{array}{cccc}
~1 & ~0 &  ~0  &  ~0 \\
~0 & ~0 &  -\, 1  &  ~ 0 \\
~0 & ~0 &  ~0  &  -\,1 \\
~0 & -\,1 &  ~ 0  &  ~0 \\
\end{array}\right] ~~~,~~~
  {\bm {\rm L}}{}^{(TM)}_{2}   ~=~
\left[\begin{array}{cccc}
~0 & ~1 &  ~0  &  ~  0 \\
~0 & ~ 0 &  ~0  &  ~ 1 \\
~0 & ~ 0 &  -\,1  &  ~ 0 \\
 ~ 1 & ~0 &  ~0  &   ~ 0 \\
\end{array}\right]  ~~~,
}    $$
\be {~~~~} { 
  {\bm {\rm L}}{}^{(TM)}_{3}   ~=~
\left[\begin{array}{cccc}
~0 & ~0 &  ~1  &  ~0 \\
~1 & ~0 &  ~ 0  &  ~ 0 \\
~0 & ~1 &  ~0  &   ~0 \\
~0 & ~0 &  ~ 0  &  -\, 1 \\
\end{array}\right] ~~~~~~,~~~
  {\bm {\rm L}}{}^{(TM)}_{4}   ~=~
\left[\begin{array}{cccc}
~0 & ~0 &  ~0  &  ~  1 \\
~0 & -\, 1 &  ~0  &  ~ 0 \\
~1 & ~ 0 &  ~0  &  ~ 0 \\
 ~0 & ~0 &  ~1  &   ~ 0 \\
\end{array}\right]  ~~~,  }
\label{tenDoF}
\ee
$\bm {VM~Adinkra~L-Matrices}$
$$ { 
  {\bm {\rm L}}{}^{(VM)}_{1}   ~=~
\left[\begin{array}{cccc}
~0 & ~1 &  ~ 0  &  ~ 0 \\
~0 & ~0 &  ~0  &  -\,1 \\
~1 & ~0 &  ~ 0  &  ~0 \\
~0 & ~0 &  -\, 1  &  ~0 \\
\end{array}\right] ~~~,~~~
  {\bm {\rm L}}{}^{(VM)}_{2}   ~=~
\left[\begin{array}{cccc}
~1 & ~ 0 &  ~0  &  ~ 0 \\
~0 & ~ 0 &  ~1  &  ~ 0 \\
 ~0 & - \, 1 &  ~0  &   ~ 0 \\
~0 & ~0 &  ~0  &  -\, 1 \\
\end{array}\right]  ~~~, }
$$
\be {~~~~} { 
  {\bm {\rm L}}{}^{(VM)}_{3}   ~=~
\left[\begin{array}{cccc}
~0 & ~0 &  ~ 0  &  ~ 1 \\
~0 & ~1 &  ~0  &   ~0 \\
~0 & ~0 &  ~ 1  &  ~0 \\
~1 & ~0 &  ~0  &  ~0 \\
\end{array}\right] ~~~~~~,~~~
  {\bm {\rm L}}{}^{(VM)}_{4}   ~=~
\left[\begin{array}{cccc}
~0 & ~0 &  ~1  &  ~ 0 \\
-\,1 & ~ 0 &  ~0  &  ~ 0 \\
 ~0 & ~0 &  ~0  &   - \, 1 \\
~0 & ~1 &  ~0  &  ~  0 \\
\end{array}\right]  ~~~. }
\label{vecDoF}
\ee

Given these sets of L-matrices, we use (\ref{GarDVs}) to find 
the corresponding ``tilde-V'' matrices with the results given respectively in 
(\ref{V-CM}),  (\ref{V-TM}), and  (\ref{V-VM})
\be \eqalign{
{\bm {\Tilde V}}{}_{12}{}^{(CM)}   ~&=~  +
{\bm {\Tilde V}}{}_{34}{}^{(CM)}   ~=~  + \, {\bm {\a^{\Hat 2}}}
   ~~~, \cr
{\bm {\Tilde V}}{}_{13}{}^{(CM)}   ~&=~   -
{\bm {\Tilde V}}{}_{24}{}^{(CM)}   ~=~     + \, {\bm {\a^{\Hat 3}}} ~~~, \cr
{\bm {\Tilde V}}{}_{14}{}^{(CM)}   ~&=~    +
{\bm {\Tilde V}}{}_{23}{}^{(CM)}   ~=~ + \,  {\bm {\a^{\Hat 1}}}   ~~~, \cr
}    \label{V-CM} \ee

\be \eqalign{
{\bm {\Tilde V}}{}_{12}{}^{(TM)}   ~&=~  -
{\bm {\Tilde V}}{}_{34}{}^{(TM)}   ~=~  + \,    {\bm {\b^{\Hat 3}}} 
 ~~~, \cr
{\bm {\Tilde V}}{}_{13}{}^{(TM)}   ~&=~  +
{\bm {\Tilde V}}{}_{24}{}^{(TM)}   ~=~  + \,    {\bm {\b^{\Hat 2}}}    ~~~, \cr
{\bm {\Tilde V}}{}_{14}{}^{(TM)}   ~&=~  - 
{\bm {\Tilde V}}{}_{23}{}^{(TM)}   ~=~   + \,    {\bm {\b^{\Hat 1}}}    ~~~, \cr
}    \label{V-TM} \ee

\be \eqalign{
{\bm {\Tilde V}}{}_{12}{}^{(VM)}   ~&=~ -
{\bm {\Tilde V}}{}_{34}{}^{(VM)}   ~=~  - {\bm {\b^{\Hat 3}}} 
  ~~~, \cr
{\bm {\Tilde V}}{}_{13}{}^{(VM)}   ~&=~   +
{\bm {\Tilde V}}{}_{24}{}^{(VM)}   ~=~ + {\bm {\b^{\Hat 2}}}   ~~~, \cr
{\bm {\Tilde V}}{}_{14}{}^{(VM)}   ~&=~  -
{\bm {\Tilde V}}{}_{23}{}^{(VM)}   ~=~   - {\bm {\b^{\Hat 1}}}   ~~~. \cr
}   \label{V-VM}
\ee
Upon comparing these equations with the forms of the equations that appear 
in (\ref{Veqz2}), one can extract the $\ell$ and $\Tilde \ell$ parameters for each 
adinkra representation.  For the $(CM)$ representation, all of the $\Tilde \ell$'s
vanish, but for $(TM)$ and $(VM)$ representations, all of the $\ell$'s vanish.  

The condition in (\ref{C1}) informs us the two adinkras corresponding to the 
BC${}_4$ adinkra representations $( {\cal R})$ and $( {\cal R}^{\prime})$ 
permit the definition of a set of Lorentzian signature Dirac Gamma matrices 
by simply making the identification
 \be  {
 {\bm \g}{}^{\mu} ~=~ {\bm k}{}^{\mu} \langle \, ({\cal R}) {\big |} ({\cal R}^{\prime}) \, \rangle
 ~~~,
} \label{gmma} \ee
from which it follows a 4D Lorentz generator is given by
\be  {
{\bm \Sigma}{}^{\mu}{}^{\nu} ~=~ i \frc 14 {\Big [} \, {\bm k}{}^{\mu} 
\langle \, ({\cal R}) {\big |} ({\cal R}^{\prime}) \, \rangle ~,~ {\bm k}{}^{\nu} 
\langle \, ({\cal R}) {\big |} ({\cal R}^{\prime}) \, \rangle  \, {\Big ]}
~~~.
} \label{gmma2} \ee
and we can express this in the form 
 \be \eqalign{ {~~~~~}
{\bm \Sigma}{}^{\mu}{}^{\nu} 
&~=~   \frc 12 \, {\bm {\Big \{ } } \,  {\big ( } \, a^{\mu}{}^{\,\Hat {\rm I}} \,  a^{\nu}{}^{\,\Hat {\rm K}} 
~-~  \ c^{\mu}{}^{\,\Hat {\rm I} \,\Hat {\rm J}} \, c^{\nu}
{}^{\,\Hat {\rm K} \,\Hat {\rm L}} \, \d{}^{\,\Hat {\rm J}}{}^{\,\Hat {\rm 
L}} \, {\big )} \, \e{}^{\,\Hat {\rm I}}{}^{\,\Hat {\rm K}}{}^{\,\Hat {\rm P}} \, {\bm \a}{}^{\Hat 
{\rm P}} \cr
&~~~~~~~~~~~+~  {\big ( } \, b^{\mu}{}^{\,\Hat {\rm I}} \,  b^{\nu}{}^{\,\Hat {\rm K}} 
~-~  \ c^{\mu}{}^{\,\Hat {\rm I} \,\Hat {\rm J}} \, c^{\nu}{}^{\,\Hat {\rm K} \,\Hat {\rm L}} 
\, \d{}^{\,\Hat {\rm J}}{}^{\,\Hat {\rm L}} \, {\big )} \, \e{}^{\,\Hat {\rm I}}{}^{\,\Hat {\rm 
K}}{}^{\,\Hat {\rm P}} \, {\bm \b}{}^{\Hat {\rm P}}   \cr
&~~~~~~~~~~~-~ i\,   {\big (}\,   a^{\mu}{}^{\,\Hat {\rm I}} \,  c^{\nu}
{}^{\,\Hat {\rm K} \,\Hat {\rm Q}} ~-~ a^{\nu}{}^{\,\Hat {\rm I}} \,  c^{\mu}{}^{\,
\Hat {\rm K} \,\Hat {\rm Q}} \, {\big )} \,  \,  \e{}^{\,\Hat {\rm I}}{}^{\,\Hat 
{\rm K}}{}^{\,\Hat {\rm P}} {\bm \a}{}^{\Hat {\rm P}}  \, {\bm \b}{}^{\Hat {\rm Q}} \cr
&~~~~~~~~~~~-~ i \,  {\big (}\,   b^{\mu}{}^{\,\Hat {\rm I}} \,  c^{\nu} {}^{\,
\Hat {\rm P} \,\Hat {\rm L}} ~-~ b^{\nu}{}^{\,\Hat {\rm I}} \,  c^{\mu}{}^{\,
\Hat {\rm P} \,\Hat {\rm L}} \, {\big )} \,  \e{}^{\,\Hat {\rm I}}{}^{\,\Hat 
{\rm L}}{}^{\,\Hat {\rm Q}} \,  {\bm \a}{}^{\Hat {\rm P}}  \, {\bm \b}{}^{\Hat {\rm Q}}   
 ~  {\Big \}  }     ~~~,  {~~~~~~~~~~~~~~}
}\label{Spn1aA}      \ee
but with the very important understanding that the coefficients $a^{\mu}
{}^{\,\Hat {\rm I}}$, $b^{\mu}{}^{\,\Hat {\rm I}}$, and $c^{\mu}{}^{\,\Hat {\rm 
I} \,\Hat {\rm J}}$ that appear in (\ref{Spn1aA}) are here strictly defined in terms
of the definitions given in (\ref{Spn1ad2}).  This operator (\ref{Spn1aA})
has the interpretation of being a spacetime spin operator constructed
from the adinkra representations (${\cal R}$) and (${\cal R}^{\prime}$). 

On the solution shown in (\ref{Spn1aa}) for the\footnote{A discussion over 
a restriction of these coefficients will be discussed in a later section.} coefficients 
$ a^{\mu}{}^{\,\Hat {\rm I}}$, $ b^{\mu}{}^{\,\Hat {\rm I}}$, and $c^{\nu} {}^{
\,\Hat {\rm K} \,\Hat {\rm L}}$ it is found that
\be  {
{\Big [} \, {\bm \Sigma}{}^{\mu}{}^{\nu} ~,~ {\bm \Sigma}{}^{\r}{}^{\s} \, {\Big ]}
~=~ - \, \eta{}^{\mu}{}^{\r} \, {\bm \Sigma}{}^{\nu}{}^{\s} ~+~  \, \eta{}^{\mu}
{}^{\s} \, {\bm \Sigma}{}^{\nu}{}^{\r} ~+~  \eta{}^{\nu}{}^{\r} \, {\bm \Sigma}
{}^{\mu}{}^{\s} ~-~  \eta{}^{\nu}{}^{\s} \, {\bm \Sigma}{}^{\mu}{}^{\r} ~~~,   }
\label{SpnOp}
\ee
which guarantees the result in (\ref{Spn1aA}) defines a spacetime spin 
operator in terms the adinkras  (${\cal R}$) and (${\cal R}^{\prime}$)
whose $\ell$ and $\Tilde \ell$ parameters determine the coefficients via 
(\ref{Spn1ad2}).  So we can pick $({\cal R})$ = $(CM)$ and $({\cal R}^{
\prime})$ = $(TM)$ or $(VM)$, but not $({\cal R}^{\prime})$ = $(CM)$, 
in order to guarantee that $\D{}_{\rI\rJ\rK\rL}^{\Hat {\rm I}  \Hat {\rm J}}$ 
does not vanish in the expressions given in (\ref{gmma}) - (\ref{SpnOp}).  

In a similar manner, a 4D chirality operator constructed from the adinkra 
representations (${\cal R}$) and (${\cal R}^{\prime}$) follows from the
expression
\be  \eqalign{  {~~~~~~~~~}
{\bm \gamma}{}^{5} ~&=~ i \frc 1{4!} \e{}_{\m \n \r \s} \, {\bm k}{}^{\mu} 
\langle \, ({\cal R}) {\big |} ({\cal R}^{\prime}) \, \rangle \, {\bm k}{}^{\nu} 
\langle \, ({\cal R}) {\big |} ({\cal R}^{\prime}) \, \rangle  \,
{\bm k}{}^{\r} 
\langle \, ({\cal R}) {\big |} ({\cal R}^{\prime}) \, \rangle \, {\bm k}{}^{\s} 
\langle \, ({\cal R}) {\big |} ({\cal R}^{\prime}) \, \rangle \cr
 ~&=~ - i \frc 1{3!} \e{}_{\m \n \r \s} \,  {\bm \Sigma}{}^{\mu}{}^{\nu}
 \,  {\bm \Sigma}{}^{\r}{}^{\s}
~~~.
} \label{gmmaFIV} \ee

Given the realization of the 4D Minkowski space Lorentz Generator on 
such a pair of adinkra representations, the Lorentz transformation properties 
on fermions in all adinkra are now well defined.  As well, the same can
be said of chirality properties.
However, we can go even beyond two such adinkras.

Consider two unordered sets of adinkras which we write (with $({\cal R})_1$
= $({\cal R})$ and $({\cal R}^{\prime})_1$ = $({\cal R}^{\prime})$)  as
\be  \eqalign{
{ \Big \{} {\cal A} {\Big  \} } ~&=~ { \Big \{} \,  ({\cal R})_1, ~
\dots ~  ({\cal R})_P \, {\Big  \} }    ~~~, \cr
{ \Big \{} {\cal B} {\Big  \} } ~&=~ { \Big \{} \,  ({\cal R}^{\prime})_1, ~
\dots ~  ({\cal R}^{\prime})_Q \, {\Big  \} }   ~~~, 
}     \ee
where $P$ and $Q$ are some integers.  Furthermore, let us assume
that any member of $\{  {\cal A} \}$ possesses the same values of all
of their $\ell$ and $\Tilde \ell$ parameters as any other member of the
set.  We also assume that any member of $\{  {\cal B} \}$ possesses the 
same values of all of their $\ell$ and $\Tilde \ell$ parameters as any 
other member of the set.  If the $a$, $b$, $c$, $A$, $B$, and $C$
coefficients are universally used to construct Dirac Gamma matrices,
they will {\em {all}} {\em {yield}} {\em {the}} {\em {same}} set of 
Dirac Gamma matrices independent of which elements from 
$\{  {\cal A} \}$ and $\{  {\cal B} \}$ are utilized.

So all adinkras with the same values of their $\ell$ and $\Tilde \ell$
parameters belong to an equivalence class with regard to their
Lorentz spacetime symmetry properties...a very satisfying result.

\section{Connecting To Riemann Surfaces}

$~~~~$ Due to the works of \cite{adnkGEO1,adnkGEO2}, we now know adinkras provide 
a periodic tessellation of Riemann surfaces with spin-structures and integer valued 
Morse divisors.  Due to these observations, the discussion given in this work provides a 
way in which the regular 4.4.4.4 tessellation of Riemann surfaces constructed from BC${
}_4$ minimal adinkras is related to spin-structures of four dimensional Minkowski space 
supermultiplets with $\cal N$ = 1 SUSY.  In other words, the adinkras have provided a 
``bridge'' between the spin-structures on a Riemann surface and spin-structures of a four 
dimensional Minkowski spacetime.

In preparation to see the connection to Riemann surfaces, we go back to the results 
presented in (\ref{AsBs}) and take the absolute values of those equations which yields 
the following
\be   \eqalign{
{\Big |}\,  {\bm \a}{}^{\,\Hat 1} \, {\Big |}  ~&=~ \left[   { {{\bm \s}{}^1}}  \,  \otimes \, { {{\bm \s}
{}^1}}  \right] ~\,~,~~ 
{\Big |}\,  {\bm \a}{}^{\,\Hat 2} \, {\Big |}  ~=~ \left[  {\bm {\rm I}}{}_2 \,  \otimes \, { {{\bm \s}
{}^1}}  \right] ~~~~,~~ 
{\Big |}\,  {\bm \a}{}^{\,\Hat 1} \, {\Big |}  ~=~ \left[  {\bm \s}{}^1 \,  \otimes \,  {\bm {\rm I}}
{}_2 \right]  ~~, \cr
{\Big |}\,  {\bm \b}{}^{\,\Hat 1}  \, {\Big |} ~&=~ \left[   { {{\bm \s}{}^1}} \,  \otimes \,   { {{\bm \s}{
}^1}}  \right]  \,~~, ~~ 
{\Big |}\,  {\bm \b}{}^{\,\Hat 2}  \, {\Big |} ~=~  \left[     { {{\bm \s}{}^1}}  \,  \otimes \,  {\bm {\rm I}}
{}_2 \right]  ~~~~, ~~ 
{\Big |}\,  {\bm \b}{}^{\,\Hat 3}  \, {\Big |} ~=~ \left[    {\bm {\rm I}}{}_2 \,  \otimes \,  { {{\bm \s}
{}^1}}   \right]  ~\,~,
} \label{AsBs2}
\ee
and, though the order is different when considering each set of matrices, $\{$ ${\big |}  {\bm \a}
{}^{\,\Hat {\rm I}} {\big |} $ $\}$ and $\{$ ${\big |}  {\bm \b}{}^{\,\Hat {\rm I}}  {\big |} $ $\}$, their 
actual elements are the same.  We may disregard the ordering and add to these three matrices 
the 4 $\times$ 4 identity element (i.\ e.\ ${\bm {\rm I}}{}_2 \,  \otimes \,   {\bm {\rm I}}{}_2$) to form 
a set we denote by $\bm {\{ {\cal V}_{(4)} \}}$ that can be written as
\be  \eqalign{ {~~~~~}
\bm {\{ {\cal V}_{(4)} \}} ~&=~ {\bm {\{ } {\bm {\cal V}}_{1} , \, {\bm {\cal V}}_{2} , \, 
{\bm {\cal V}}_{3} , \,  {\bm {\cal V}}_{4} {\bm  \}} }  ~=~ 
{\bm \{ }  ~  {\bm {\rm I}}{}_{2 \times 2} \otimes   
{\bm {\rm I}}{}_{2 \times 2} , ~  {\bm {\rm I}}{}_{2 \times 2}  \otimes  {\bm \s}^1, ~ 
{\bm \s}^1 \otimes   {\bm {\rm I}}{}_{2 \times 2} , ~  {\bm \s}^1 \otimes
{\bm \s}^1 ~  {\bm \} }  ~~~,
} \label{Cgrp4z}
\ee
which alternately can also be written in the form of
\be  \eqalign{
\bm {\{ {\cal V}_{(4)} \}}  ~=~&{\bm \{} (), ~ (12)(34), ~ (13)(24), ~ (14)(23) {\bm \} }   
  ~~~,  \cr
} \label{Cgrp4}
\ee
by use of cycle notation.  These matrices form a group, the Klein Vierergruppe, but they
do not satisfy the conditions in (\ref{GarDNAlg2}).  In order to achieve this, L-matrices
\cite{permutadnk}
\begin{equation}
 {\bm \rL} ~=~ 
     {\bm {\cal S}} \,{\bm  \cdot} \, {\bm {\cal P}}
\label{aas0}
\end{equation}
which are signed permutations must be introduced.  The factors ${\bm {\cal P}}$ denote
special quartets of permutations (as identified in the work of \cite{permutadnk}) acting 
on four objects and the factors ${\bm {\cal S}}$ are 4 $\times$ 4 matrices given the 
nomenclature of ``Boolean Factors'' in this work.  These have the forms indicated
by
\be 
(p_1 2^0 + p_2 2^1 + p_3 2^2 + p_4 2^3)_b ~\equiv~   \left( \begin{array}{cccc}
				(-1)^{p_1} & 0 & 0 & 0 \\
				0 & (-1)^{p_2} & 0 & 0 \\
				0 & 0 & (-1)^{p_3} & 0 \\
				0 & 0 & 0 & (-1)^{p_4}  \end{array}
	\right)~~~,
\ee
where $p_1$, $p_2$, $p_3$, and $p_4$ are bits taking on values of either one 
or zero.  There are sixteen ``even'' Boolean Factor quartets that may be combined 
with the elements of $\bm {\{ {\cal V}_{(4)} \}}$ to form the L-matrices indicated in 
(\ref{aas0}). These are,
\be
\eqalign{
{~~~~}
{{\bm S}_{ {\cal V}_{(4)} }} [\a]: &  
\{ (12)_b , (10)_b , (0)_b , (6)_b \} , 
\{ (6)_b , (12)_b , (0)_b , (10)_b  \} , 
\{  (14)_b , (8)_b , (2)_b , (4)_b  \} , {~~~~~~~~~~~~~~~~~~~~} \cr
& \{ (4)_b , (14)_b , (2)_b , (8)_b  \} , 
\{ (8)_b , (14)_b , (4)_b , (2)_b  \} , 
\{ (2)_b , (8)_b  , (4)_b , (14)_b \} , \cr
&\{(10)_b , (12)_b , (6)_b , (0)_b \} , 
\{  (0)_b , (10)_b , (6)_b , (12)_b  \} , 
\{  (14)_b , (4)_b , (8)_b , (2)_b  \} , \cr
& \{  (4)_b , (2)_b , (8)_b , (14)_b \} , 
\{  (12)_b , (6)_b , (10)_b , (0)_b  \} , 
\{  (6)_b , (0)_b , (10)_b , (12)_b  \} , \cr
& \{ (10)_b , (0)_b , (12)_b , (6)_b  \} , 
\{  (0)_b , (6)_b , (12)_b , (10)_b  \} ,
\{  (8)_b , (2)_b , (14)_b , (4)_b  \} ,  \cr 
 &\{ (2)_b , (4)_b , (14)_b , (8)_b  \}  ~~. {~~~~~~}
 }  \label{e:CMEvenV}
 \ee
Though only even Boolean Factors are listed above,  ``odd''  quartets
exist.  Given a Boolean Factor 
quartet of the form $\{ (p)_b , (q)_b , (r)_b , (s)_b \}$, its
 ``antipodal antonym'' \cite{antonK} $\{ (P)_b , (Q)_b , (R)_b , (S)_b \}$ via
\be
\{ (P)_b , (Q)_b , (R)_b , (S)_b \} ~=~
 \{ (15 \, - \, p)_b , (15 \, - \, q)_b , (15 \, - \, r)_b , (15 \, - \, s)_b \} ~~~.
 \label{BF}
 \ee
If $\{ (p)_b , (q)_b , (r)_b , (s)_b \}$ together 
with a specific quartet of permutations ${\bm {\cal P}}$ form a set of L-matrices,
this will remain true if any number of the Boolean Factors within the quartet are 
replaced by their antonyms.  An explicit example shows how this 
works.
 We can form a set of L-matrices by forming an inner product using
\be \eqalign{
{\bm \rL} ~&=~  {\bm \{} (12)_b\, , (10)_b \, , (0)_b \, , (6)_b {\bm \}} \, {\rm \cdot}  
\, {\bm \{} (), ~ (12)(34), ~ (13)(24), ~ (14)(23) {\bm \} }  \cr
~&=~  {\bm \{} (12)_b (), \, (10)_b (12)(34), \, (0)_b (13)(24), \, (6)_b
(14)(23) {\bm \}}  ~~~,
} \ee
which can be shown to satisfy the conditions in (\ref{GarDNAlg2}).  The antipodal 
antonym L-matrices are obtained from  
\be \eqalign{ {~~}
{\rm {antipodal~antonym}} \left[ {\bm \rL} \right] ~&=~  {\bm \{} (3)_b , (5)_b , (15)_b , 
(9)_b {\bm \}} \, {\rm \cdot}  
\, {\bm \{} (), ~ (12)(34), ~ (13)(24), ~ (14)(23) {\bm \} }  \cr
~&=~  {\bm \{} (3)_b (), \, (5)_b (12)(34), \, (15)_b (13)(24), \, (9)_b
(14)(23) {\bm \}} ~=~ -\, {\bm \rL} ~~~.
} \ee
Simple calculations show these L-matrices satisfy the conditions in (\ref{GarDNAlg2}).  
The antipodal antonym corresponds to a sign exchange of a single fermionic node.
By performing a simultaneous sign change on {\em {all}} the boson nodes or {\em 
{all}} the fermion nodes, one can eliminate these as independent representations.
In terms of the field theory described by the adinkras, this is simply a fermionic
field redefinition by a minus sign.  
 
We partition the sixteen Boolean Factors according to
\be
{{\bm S}_{ {\cal V}_{(4)} }} [\a] ~=~ {{\bm S}_{ {\cal V}_{(4)} }} [{\a}^-] ~ \cup ~
{{\bm S}_{ {\cal V}_{(4)} }}[{\a}^+]
\ee
where $\a^-$ = 1, 3, 5, 7, 10, 12, 14, and 16 while  $\a^+$ = 2, 4, 6, 8, 9, 11, 
13, and 15.  The $\ell$ coefficients vanish for all members of ${{\bm S}_{ {\cal 
V}_{(4)} }} [{\a}^-] $ while the  $\Tilde \ell$ coefficients vanish for all members 
of ${{\bm S}_{ {\cal V}_{(4)} }} [{\a}^+] $.  From discussion in previous 
chapters, any choice of spin structure on the Riemann surface that includes 
one factor from the $\a^-$ set and one from the $\a^+$ set should lead to 
a spacetime spin generator among minimal 4D, $\cal N$ = 1 supermultiplets.

We are now in position to make contact with the work in prior papers 
\cite{adnkGEO1,adnkGEO2}.  For our purposes, the most relevant points 
from these works can be captured in one formula and one
figure.  It was noted when adinkras are utilized to
reconstruct Riemann surfaces, there arises a relation between the genus
of the Riemann surface $g$, the number of distinct colors $N$ of the links
according to the formula,
\be
g ~=~ 1 ~+~ {\rm d} \, \left( \,  N~-~ 4   \, \right) ~~~,~~~
\label{genus}
\ee
where the number of closed nodes (also equal to the number of open nodes) 
is d.  Since all adinkras associated with BC${}_4$ correspond to d = 4 and possess 
exactly four colors,  the second factor in the sum in (\ref{genus}) vanishes. 
For the special case of $N$ = 4, all adinkras, even ones not associated with
BC${}_4$, ``live'' on the torus independent
of the number of nodes. 

The formula 
in (\ref{genus}) informs us the any BC${}_4$ adinkra, such as the chiral 
supermultiplet shown in Fig.\ \# 1,  can be used as a tessellation of a torus.  
In Fig.\ \#\ 4, the chiral supermultplet adinkra has been ``cut open,'' ``laid flat,'' 
``replicated,''  ``glued together,'' and used for a 4.4.4.4 tiling of a torus.
$$
\vCent
{\setlength{\unitlength}{1mm}
\begin{picture}(-20,0)
\put(-70,-240){\includegraphics[width=8.0in]{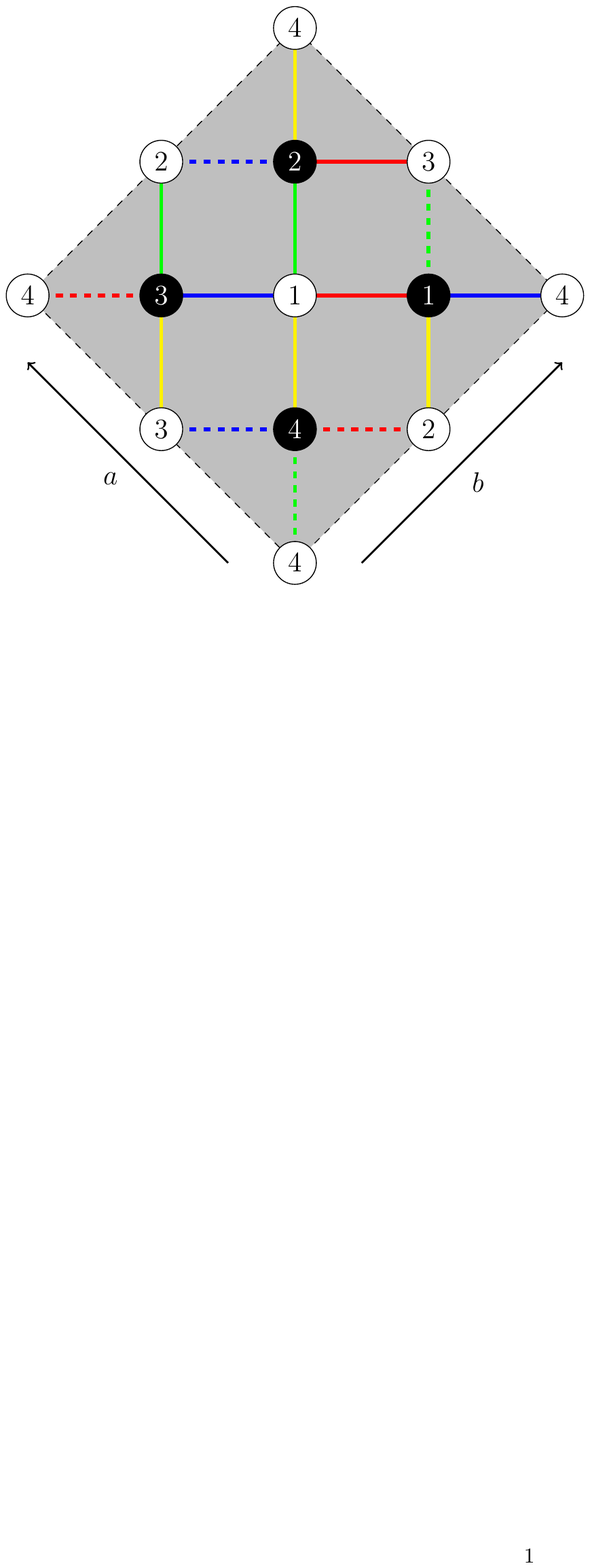}}
\end{picture}}
\nonumber
$$
\vskip3.3in
$ ~~~~~~~~~~~~~~{\bm {\rm {Figure ~ 4:~}}} 
\bm {\rm Chiral~Supermultiplet~Adinkra~On~Torus~}
$ \vskip0.1in \noindent
The image in Fig.\# 4 also includes the a-cycle and b-cycle of the torus.  Though
we only show the adinkra from Fig.\ \# 1, the adinkras from Fig.\ \# 2 and
Fig.\ \# 3, can be treated in the same manner.

Forgetting about the dashing on the adinkras, the monodromy data associated 
to the Riemann surfaces in \cite{adnkGEO1} leads to a set of matrices of the 
type shown here in (\ref{Cgrp4z}) and (\ref{Cgrp4}). The authors of \cite{adnkGEO2} 
construct a dictionary between odd dashings and certain spin structures on the 
adinkra Riemann surface. The Boolean factors used above encode odd dashings 
on the adinkra.  By combining these two dictionaries, the spin operators constructed 
from adinkras for the four dimensional minimal supermultiplets can be built from 
pairs of (non isomorphic!) spin structures on these Riemann surfaces.

When these are clarified, it will connect Riemann surfaces with adinkra tessellation 
to the spin operators constructed from 
adinkras for the four dimensional minimal supermultiplets.  The fermionic holoraumy 
matrices ${\bm {\Tilde V}}$ ${}_{\rI\rJ}^{(\cal R)}$ of (\ref{GarDVs}) provide ``colored and 
decorated'' plaquettes as shown in Fig.\ \# 4, equipped with an ``isospin R-symmetry,'' 
used for the tessellation of the Reimann surfaces and thus are the origins of the 
spacetime chirality and Lorentz symmetries for the 4D, $\cal N$ = 1 supermultiplets 
via a ``spin from iso-spin'' approach.


 \noindent
{\bf Acknowledgements}\\[.1in] \indent
We would like to thank Charles Doran, and Jordan Kostiuk for very productive
discussions and suggestions while writing this paper.  We further acknowledge
J.K. for the image of the chiral supermultiplet on the torus.
This work was partially supported by the National Science Foundation 
grant PHY-1315155 and also supported by the Maryland Center for
String and Particle Theory.    SJG also acknowledges the generous 
support of the Provostial Visiting Professorship Program and the 
Department of Physics at Brown University for the very congenial and 
generous hospitality during the period of this work.  Finally, we
acknowledge the encouraging and stimulating atmosphere that was 
created by all participants in the 2016 ``Brown University Adinkra Math/Phys 
Hangout'' (19-23 Dec. 2016).

\end{document}